# On-chip lateral Si:Te PIN photodiodes for room-temperature detection in the telecom optical wavelength bands

Mohd Saif Shaikh[1,2], Shuyu Wen[1,3], Mircea-Traian Catuneanu[2], Mao Wang[4*], Artur Erbe[1,2], Slawomir Prucnal[1], Lars Rebohle[1], Shengqiang Zhou[1], Kambiz Jamshidi[2], Manfred Helm[1,2], Yonder Berencén[1**]

[1]Helmholtz-Zentrum Dresden-Rossendorf, Institute of Ion Beam Physics and Materials Research, Bautzner Landstrasse 400, 01328 Dresden, Germany
[2]Dresden University of Technology, 01062 Dresden, Germany
[3]Institute of Semiconductors, Chinese Academy of Sciences, Bejing, China
[4]Laboratory of Micro-Nano Optics, College of Physics and Electronic Engineering, Sichuan Normal University, Chengdu 610101, Peoples Republic of China

Corresponding author: m.wang@hzdr.de*; y.berencen@hzdr.de**

## Abstract

Photonic integrated circuits require photodetectors that operate at room temperature with sensitivity at telecom wavelengths and are suitable for integration with planar complementary-metal-oxide-semiconductor (CMOS) technology. Silicon hyperdoped with deep-level impurities is a promising material for silicon infrared detectors because of its strong room-temperature photoresponse in the short-wavelength infrared region caused by the creation of an impurity band within the silicon band gap. In this work, we present the first experimental demonstration of lateral Te-hyperdoped Si PIN photodetectors operating at room temperature in the optical telecom bands. We provide a detailed description of the fabrication process, working principle, and performance of the photodiodes, including their key figure of merits. Our results are promising for the integration of active and passive photonic elements on a single Si chip, leveraging the advantages of planar CMOS technology.

## 1. Introduction

All-Si optoelectronic integrated circuits hold great promise for future applications in telecommunication, integrated optics, microelectronics, and sensing. Efficient light sources and passive photonic components for the O (1.3 $\mu$m) and C (1.5 $\mu$m) bands have been extensively researched, but the development of cost-effective, room-temperature photodetectors in these spectral ranges remains a challenging task [1–4]. Currently, Ge and InGaAs are the materials of choice for photodetection in this range due to their high absorption coefficients, but they require heterogeneous integration on Si substrates which can be complex and costly [5–7]. Alternatively, the monolithic integration offers a simpler and more scalable approach that involves fabricating both the Si photodetector and other integrated Si components on the Si substrate using standard semiconductor processing techniques. This approach circumvents the need for complex assembly and alignment processes. Moreover, it offers good compatibility with complementary metal-oxide-semiconductor (CMOS) technology, enabling the integration of photodetectors with electronic circuits on the same Si chip at a lower cost. However, the performance of monolithically integrated silicon photodetectors at the telecom bands is limited by the properties of intrinsic Si, which has a low absorption coefficient ($10^{-8}$ cm$^{-1}$) in the aforementioned spectral ranges. To address this limitation, the intentional introduction of deep-level impurities like transition metals (Ti, Au, Ag) and chalcogens (S, Se, and Te) at concentrations above the solid-solubility limit has been proposed to create an impurity band within the Si bandgap[8–15]. This leads to room-temperature absorption in the O- and C-telecom bands (0.9 eV, 1.3 µm and 0.8 eV, 1.5 µm), making it a promising approach for Si-based photodetection. Hyperdoped or supersaturated Si is a new type of Si created through this approach.

The fabrication of PN photodiodes is widely used to evaluate the potential of hyperdoped silicon for infrared applications[16–20]. Hyperdoped Si is typically used as an absorption medium for photons with lower energy and simultaneously as an n-type doped region that is deliberately created atop a p-type doped silicon substrate to form the PN junction. Several experimental demonstrations leading to room-temperature vertical hyperdoped Si PN photodetectors with an extended photoresponse up to 5.9 µm have been reported[16–20]. Yet, they may not be compatible with passive photonic components like waveguides, attenuators, and modulators that are required for monolithic integration on a very large scale. To address this, a conceptual planar design of an array of lateral PIN photodiodes based on hyperdoped Si has been proposed, which is compatible with the VLSI Si technology [19].

In this work, we present the first experimental demonstration of a lateral PIN photodiode with Te-hyperdoped Si as the active absorber material, which is operated at room temperature for O- and C-telecom bands. Our approach involves the formation of shallow n-type (Te-hyperdoped) and p-type (B-doped) regions by ion implantation followed by recrystallization through a single laser annealing pulse. We investigate the effect of the Te doping concentration on the device performance by fabricating PIN Si:Te devices with two concentrations (Te-0.5 at.% and Te-1.5 at.%). Compared to the Te-0.5% devices, the Te-1.5% devices exhibit a significant increase in responsivity, with maximum values of $1 \times 10^{-3}$ A/W and $4 \times 10^{-4}$ A/W at wavelengths of 1300 nm and 1500 nm, respectively. However, the Te-0.5% devices show a lower noise equivalent power (NEP) value of about $2 \times 10^{-7}$ W/√(Hz) due to fewer material defects. Additionally, we evaluated the frequency response characteristics and measured a 3dB cut-off frequency of 268 kHz with an applied bias of -1V.

2. **Experimental details**

In this study, an intrinsic Si (100) wafer was used as the substrate, with a resistivity of >10 kΩ•cm and a thickness of 380 µm. UV photolithography and ion implantation were utilized to form p-type and n-type regions on the substrate, laterally doped with B and Te to a depth of around 100 nm, respectively, using an interdigitated structure, as shown in Fig. 1(a-b). The dimensions of the device, not to scale, are illustrated in Fig. 1(b). After ion implantation, a recrystallization and dopant activation process was performed using a single pulse of a XeCl laser with an energy of 1.1 Jcm$^{-2}$ for both the B- and Te-doped regions. Raman spectroscopy was employed to investigate the microstructural properties of the devices during the synthesis and upon the pulsed laser annealing (PLA). The measurements were conducted at room temperature in a backscattering configuration from 100 cm$^{-1}$ to 600 cm$^{-1}$ with a resolution of around 3 cm$^{-1}$ under a laser excitation of 532 nm.

Two PIN devices were fabricated with 0.5 and 1.5 atomic % concentrations of Te, all other parameters were kept the same as detailed above. Subsequently, 200-nm thick Al and 40nm/160nm Ti/Au were deposited by thermal and e-beam evaporation to form the metal contacts to the p- and n-type regions, respectively. Surface passivation or anti-reflective coating was not applied. The devices were then wire-bonded on ceramic chip carriers to enable optoelectronic measurements while illuminating the devices from the back side (top illumination would cause shadowing of the Te-hyperdoped sub-surface layer, as evident in Fig. 3(a)). A home-built photodetector characterization setup embedded in an optical enclosure was used to perform a comprehensive characterization of the photodetectors. The spectral responsivity was measured using a monochromator (Horiba Triax 550) and a Quartz Tungsten Halogen (QTH) lamp coupled to a home-built microscope with broadband and multimode optical fiber in the range of 400 - 2200 nm. A chopper was employed to modulate the impinging light on the photodetectors after passing through an optical long-pass filter of 1200 nm (FELH1200), which eliminates second-order optical contributions from the grating. The generated photocurrent from the photodetector was coupled to a low-noise current preamplifier (SR570) and subsequently to the lock-in-amplifier (SRS830) for signal acquisition.

A calibrated pyroelectric detector (Newport DET-L-PYK5-R-P) was employed to evaluate the optical response of the entire system, as these detectors tend to have almost a flat spectral infrared response from 0.8 μm to the far-infrared region (FIR). Dark current-voltage (IV) measurements were performed using a Semiconductor Device Analyzer (Agilent B1500A). Instead of estimating the NEP theoretically for the Si:Te PIN detectors, the noise current spectral density was measured, and the minimum signal power that produces an SNR=1 was extracted[21]. We performed these measurements using a modulated laser of 1550 nm at the frequency of interest and passing this light through a set of neutral density (ND) filters and performing the signal acquisition on a spectrum analyzer (Agilent Technologies PXA N9030A signal Analyzer (3 Hz – 50 GHz)). For the verification of the measurement procedure, a commercially calibrated detector (Thorlabs FDG03-CAL Cert. no. 21229134757) was used. More details on the measurement procedures can be found in the supporting information.

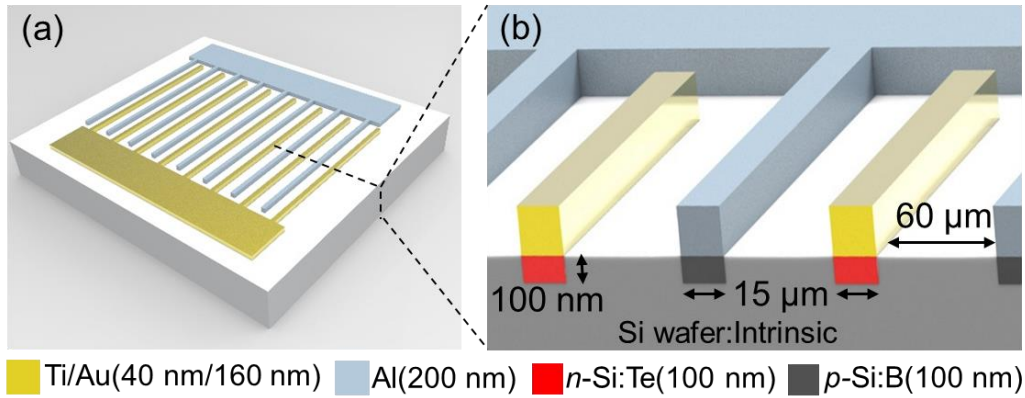

Figure1. PIN Si:Te photodetector a) top view of the lateral Si:Te PIN photodiodes, b) cross-section of the PIN photodetector along the dotted line in Fig.1 (a) showing the sub-surface doped *p* and hyperdoped *n* regions with the respective dimensions.

## 3. Results and discussion

### 3.1 Material synthesis and analysis

The implantation profile of Te and B dopants was simulated using Stopping and Ranging of Ions in Materials (SRIM) software, and the resulting profiles are shown in Fig. 2 (a) and (b). The implantation depth and concentration of each dopant were controlled by the implantation energy and fluence, respectively. Two sets of Si:Te PIN devices were fabricated, each following a two-step ion implantation to ensure a nearly flat doping profile. The first set of devices contained a Te doping concentration of $2.5 \times 10^{20}$ cm$^{-3}$ (0.5 at.%) and the second set a Te concentration of $7.5 \times 10^{20}$ cm$^{-3}$ (1.5 at.%) at the fluence and energy given in the inset of Fig. 2(a). P-type doping was performed by implanting B atoms at a concentration of $1 \times 10^{18}$ cm$^{-3}$ for both devices following a two-step implantation. The estimated doped layer thickness for both Te and B atoms was around 100 nm.

Raman spectra of the hyperdoped Si substrate were measured before and after PLA for the Te and B implanted regions along with a single-crystalline pristine Si (Fig. 2 (c) and (d)). The second-order transverse acoustic phonon (2TA) mode at 303 cm$^{-1}$ and the dominant transverse optical (TO) mode at 520 cm$^{-1}$ were used as reference modes to inspect the crystalline quality of the hyperdoped Si upon the annealing treatment. After Te implantation, for both Te concentrations (Te-0.5% and Te-1.5%), the TO mode of Si almost vanished due to amorphization originated by the high fluence implantation. Additionally, Te, being heavier than B, induced significant amorphization characterized by broad Raman bands centered at around 150 cm$^{-1}$ and 467 cm$^{-1}$. The Si TO peak remained visible for both Te concentrations due to the underlying crystalline Si substrate, although it was slightly higher for Te-0.5%. On the other hand, we observed no amorphization-related bands induced by the implantation in the B implanted region.

After PLA, the TO mode of Si increased again in intensity due to the recovery/recrystallization of the implanted regions. Due to the dopant activation by PLA, we observed an asymmetric broadening of the TO mode of Si, which accounts for the Fano-type electron-phonon interaction. This asymmetric broadening to the left side for n-type dopant is well known[22–25] and is analyzed further in the supporting information (see Fig. S3). We found a more pronounced asymmetric broadening for a sample with a higher Te concentration after PLA, which is attributed to a higher free electron concentration.

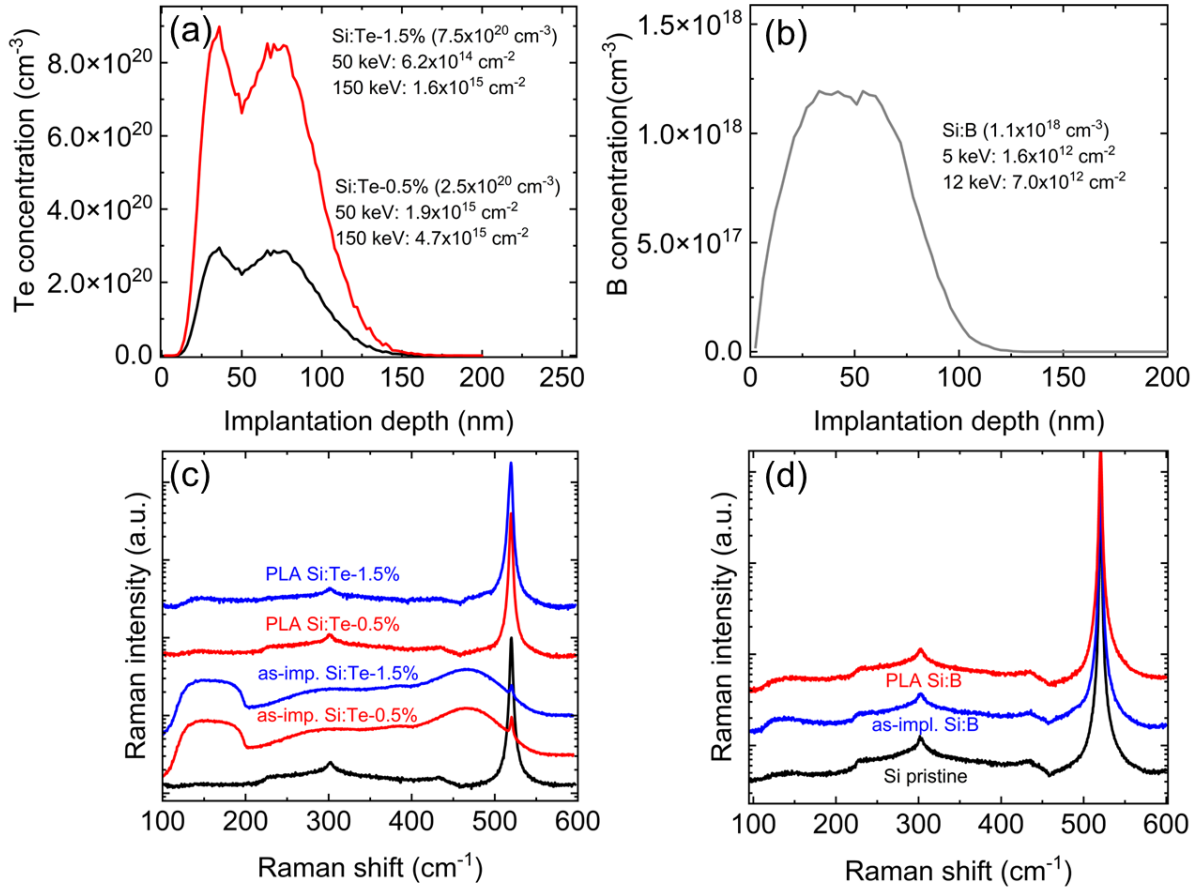

Figure 2. SRIM simulation of a) Te implantation profile in Si and b) B implantation profile in Si. Raman spectra before and after PLA of c) Te implanted Si and d) B implanted Si.

The electrical properties of the hyperdoped Si samples were characterized by room-temperature Hall-effect measurements. For the Te-doped samples, we measured an active electron concentration of $8.5 \times 10^{19}$ cm$^{-3}$ ($4.3 \times 10^{20}$) and electron mobility of 52 cm$^2$/Vs (37 cm$^2$/Vs) for the Te-0.5% (Te-1.5%) samples, respectively. The electron mobility decreases with increasing Te concentration due to the increase in the density of ionized impurities and the Coulomb scattering from the charged impurities. For the B-doped region, we measured a hole concentration of $4.3 \times 10^{17}$ cm$^{-3}$ and a hole mobility of $1.2 \times 10^4$ cm$^2$/Vs.

*3.2 Dark current-voltage (IV) and telecom-wavelength photoresponse*

The room-temperature current-voltage (IV) characteristics of lateral Si:Te PIN detectors were measured under dark conditions, and we observed a diode behavior with a rectification ratio of around 230 for Si:Te-0.5% and around 50 for Si:Te-1.5% detectors measured at 1 V and -1 V (see Fig. 3(b)). We fitted each IV curve with a single-diode model using the equation

$$I = I_0 \left[ e^{\frac{q(V-IR_S)}{\eta k_B T}} - 1 \right] + \frac{V - IR_S}{R_{shunt}} \quad (1),$$

where $V$ and $I$ are the voltage and current, $R_s$ and $R_{shunt}$ are the series and parallel resistances, respectively, $I_0$ is the saturation current, $q$ is the electron charge, $\eta$ is the ideality factor, $k_B$ is the Boltzmann constant and $T$ is the temperature.

The parameters of the fitting are summarized in table 1, along with the reference Si:Te (with Te-2.0%) PN vertical device[20]. Our lateral Si:Te PIN devices exhibit a better rectification ratio, lower dark current, and lower saturation current, although we observed an increased ideality factor. The Si:Te-1.5% device showed a higher dark current, likely due to the presence of more defects associated with a higher Te concentration implantation. This is consistent with the broad absorption band observed at around 467 cm$^{-1}$ in the Raman experiments (see Fig. S3 in the supplementary information).

Table 1. Comparison of parameters acquired from IV characteristics.

| Parameters | Si:Te PIN lateral det. 0.5% at. | Si:Te PIN lateral det. 1.5% at. | Ref. Si:Te PN vertical det. 2.0% at.[20] |
|---|---|---|---|
| Series resistance ($R_s$) | 8.1 kΩ | 20.3 kΩ | 6.4 Ω |
| Shunt resistance ($R_{sh}$) | 3.1 MΩ | 96.2 kΩ | 850.0 Ω |
| Saturation current ($I_0$) | 0.3 µA | 5.3 µA | 68.0 µA |
| Ideality factor ($\eta$) | 2.3 | 4.8 | 2.2 |
| Rectification ratio | 230 | 50 | 36 |
| Dark current at 0V (-1V) | 3.9 nA (0.4 µA) | 36.0 nA (0.1 mA) | 10.0 µA (1 mA) |

Next, we measured the room-temperature photoresponse of the Si:Te PIN detectors in response to a 1550 nm laser at different input light powers (Fig. 3 (c)). We found that both detectors show a linear response (exponent α=1 shown in the inset Fig. 3(c)) to the incident optical power from 0.7 µW to 0.7 mW at -1V and at 0V applied bias. Upon biasing (-1V) and for the same input laser power, we observed a remarkable improvement of two and four orders of magnitude in the generated photocurrent for Te-0.5% and Te-1.5% respectively. This will be discussed later.

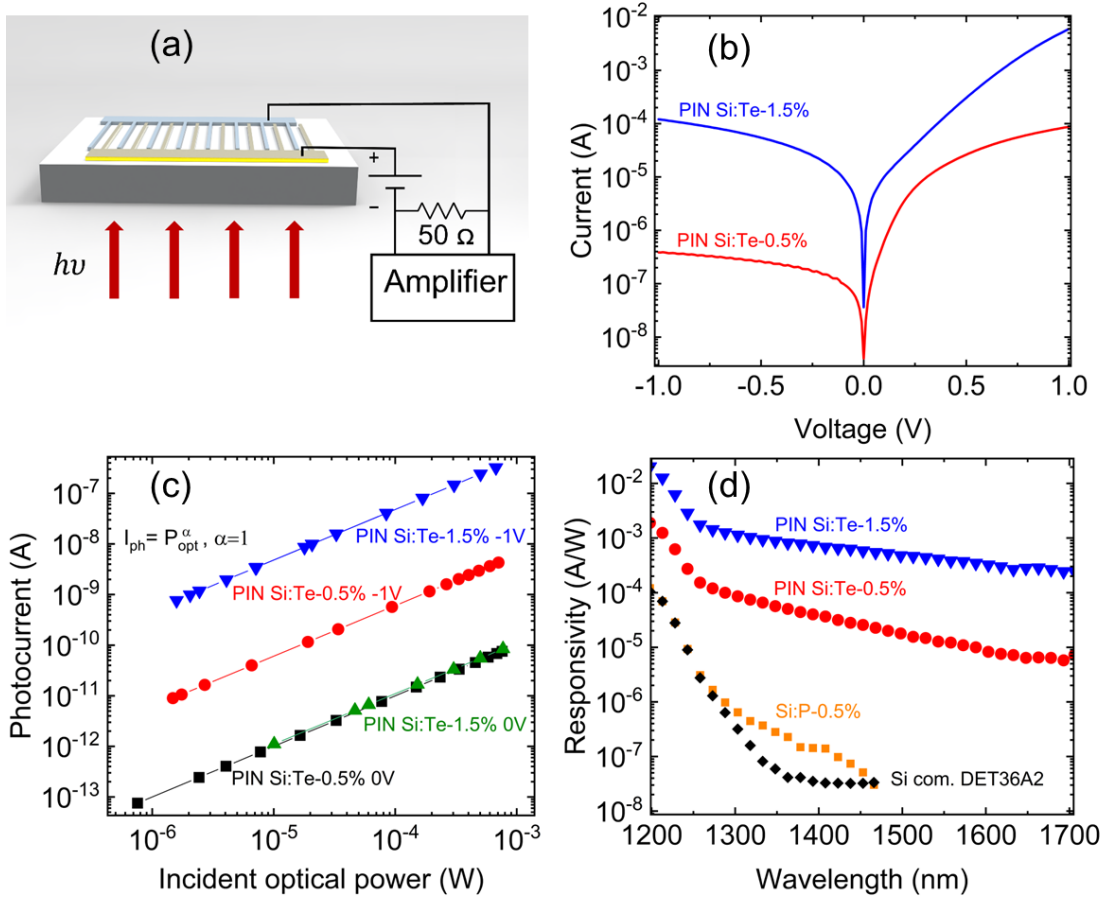

Figure 3. PIN Si:Te detector with 0.5% and 1.5% Te concentration, a) schematic of device representing rear-side illumination and biasing circuit b) Dark IV characteristics c) Photocurrent versus input optical power at a wavelength of 1,550 nm for both devices at 0V and -1V reverse bias, d) spectral responsivity under -1V reverse bias.

We also investigated the spectral responsivity of Si:Te PIN detectors under a reverse bias of -1V, along with a commercial Si detector (Thorlabs-DET36A2) and a phosphorus n-type-hyperdoped Si control device (Fig. 3 (d)). Our results revealed that the Si commercial detector reached the noise floor at around 1350 nm, while the Si:P control device exhibited a kink in responsivity from 1300 to 1450 nm, indicating generated photocurrent from defects created during implantation and PLA, with a maximum of $4\times10^{-7}$ A/W at around 1330 nm. This photocurrent might be originating from the defects created during the implantation and PLA *e.g.* Si-divacancy defects[26,27,16,28]. Interestingly, the Si:Te PIN detectors exhibit a different profile and an overall higher and extended responsivity than the control device, indicating the existence of a Te-related intermediate band within the Si bandgap enabling the absorption of photons in the telecom optical bands. We observed that the responsivity and external quantum efficiency (EQE) of the Si:Te PIN detectors increase with increasing Te concentration, and the devices with Te-1.5% exhibit orders of magnitude higher EQE values than the state-of-the-art of vertical Si:Te PN detector [20]. The results are summarized in Table 2, which shows the responsivity and calculated EQE of Si:Te PIN detectors with different Te concentrations at 1300 nm and 1500 nm wavelengths under the applied bias of 0V and -1V.

Table 2. Summary of responsivity and calculated EQE at applied bias.

| Parameters | Wavelength (nm) | Si:Te PIN det. 0.5% at. | | Si:Te PIN det. 1.5% at. | |
|---|---|---|---|---|---|
| | | 0V | -1V | 0V | -1V |
| Responsivity (A/W) | 1300 | 5.3x10$^{-7}$ | 8.7x10$^{-5}$ | 7.6x10$^{-7}$ | 1.1x10$^{-3}$ |
| | 1500 | 1.0x10$^{-7}$ | 1.2x10$^{-5}$ | 1.1x10$^{-7}$ | 4.2x10$^{-4}$ |
| EQE (%) | 1300 | 7.1x10$^{-5}$ | 8.3x10$^{-3}$ | 7.2x10$^{-5}$ | 1.0x10$^{-1}$ |
| | 1500 | 7.0x10$^{-6}$ | 1.0x10$^{-3}$ | 1.2x10$^{-5}$ | 3.0x10$^{-2}$ |

## *3.3 Noise Equivalent Power, Detectivity and Frequency response*

We also determine the sensitivity of our photodetectors, expressed as noise equivalent power NEP($\lambda, f$) (See Suppl. Info), defined as the minimum optical power required to achieve a signal-to-noise ratio (SNR) of 1 at the frequency of interest. A lower NEP value indicates a higher sensitivity of the photodetector. To measure NEP, we used a 1550 nm laser, modulated at a frequency of 70 Hz, and a spectrum analyzer with an electrical bandwidth set to 1Hz. We have obtained experimental data for devices with two Si:Te concentrations under 0V and -1V bias, as well as for a commercially calibrated Ge detector (FDG03-CAL) which also validates our experimental methodology (Fig. 4 a). We found the NEP$_{1550\,nm, 70\,Hz}$ of the calibrated Ge detector to be (1.3x10$^{-12}$ W/$\sqrt{(Hz)}$) in good agreement with the value (1.0x10$^{-12}$ W/$\sqrt{(Hz)}$) in the datasheet. Likewise, we determined the NEP values for our Si:Te PIN photodetectors. Interpolating the experimental data, which are fitted with a slope of 1 to the SNR value of 1. For the Si:Te-0.5% PIN detector under zero bias, we deduced the NEP$_{1550\,nm, 70\,Hz}$ to be 2.4x10$^{-7}$ W/$\sqrt{Hz}$. This value is two orders of magnitude lower than that of Si:Te-1.5% PIN detector, which has a higher dark current (Fig. 3 b). Upon reverse biasing at -1V, we found the NEP$_{1550\,nm, 70\,Hz}$ of Si:Te-1.5% PIN detector to be 5.8x10$^{-7}$ W/$\sqrt{Hz}$, and for Si:Te-0.5% PIN detector 7.8x10$^{-7}$ W/$\sqrt{Hz}$. NEP is the ratio of the noise current $i_n$ (in A), and the current responsivity $R_i$ (in A/W). Although the Si:Te-1.5% PIN device has one order higher responsivity compared to Si:Te-0.5% the NEP value is high because of the higher noise current which originates from the inferior material quality due to higher induced defects with increased doping.

The NEP depends on the square root of the detector area ($A$) as well as the square root of the measurement bandwidth ($\Delta f$) as both parameters affect the noise of the detector. Therefore, to fairly compare different detectors, a key figure of merit is the specific detectivity which normalizes NEP to 1 Hz noise bandwidth and the area of the detector to 1 cm$^2$ [29]. Specific detectivity is given by $D^* = \frac{\sqrt{A\Delta f}}{NEP}$, where $\Delta f$ is the electrical bandwidth (1 Hz) and, $A$ is the detector's active area, which is 0.09 cm$^2$ for PIN Si:Te devices and 0.07 cm$^2$ for the commercial Ge detector. For the Si:Te PIN detectors we found a maximum specific detectivity of 1.2x10$^6$ cmHz$^{1/2}$ W$^{-1}$ for the Si:Te-

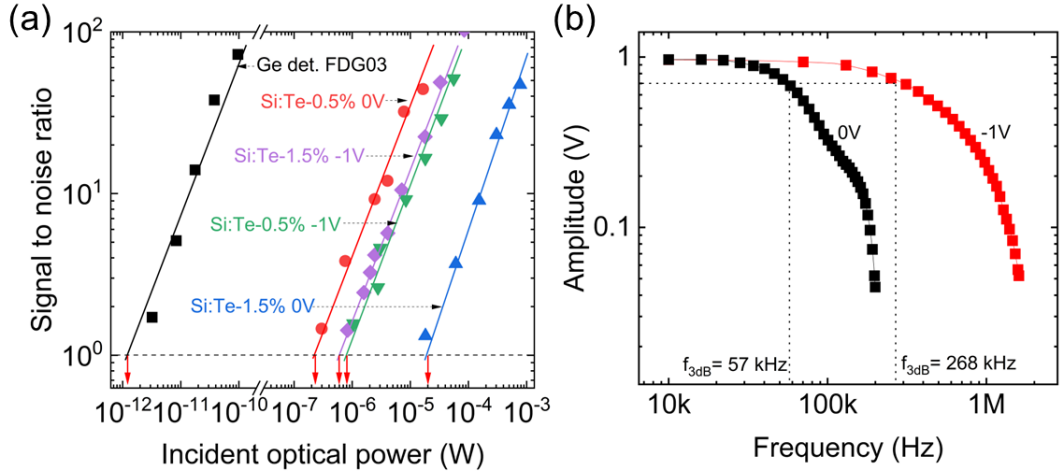

Figure 4. a) Noise equivalent power measurement of commercial Ge detector FDG03 and Si:Te PIN detectors, b) Frequency response of Si:Te PIN detector with 0V and -1V reverse bias.

0.5% PIN detector when unbiased. The measured NEP and calculated specific detectivity values of all the devices are summarized in table 3.

In addition to measuring the sensitivity of the photodetectors, we also evaluated their frequency response. The 3dB cut-off frequency is a measure of the frequency response of a detector, which is defined as the frequency at which the detector's output signal is reduced to half of its maximum amplitude. In the case of the Si:Te PIN detectors, the 3dB cut-off frequency was found to be 57 kHz when the device was unbiased and 268 kHz when biased, as shown in Fig. 4(b). The interdigitated structures of the devices cover a large area which results in a large response time (RC constant) that limits the device response. However, applying bias increases the area of the depletion region, reduces the capacitance, and increases the 3dB cut-off frequency.

Table 3. Measured NEP and specific detectivity for PIN Si:Te and commercial calibrated Ge detectors.

| measured at 300K | | NEP$_{1550\,nm, 70\,Hz}$ (W/√($Hz$)) | Specific detectivity (cmHz$^{1/2}$ W$^{-1}$) |
|---|---|---|---|
| FDG03-CAL | Datasheet | 1.0x10$^{-12}$ | ------ |
|  | Measurement | 1.3x10$^{-12}$ | 0.2x10$^{12}$ |
| Si:Te-0.5% | 0V | 2.4x10$^{-7}$ | 1.2x10$^6$ |
|  | -1V | 7.8x10$^{-7}$ | 3.8x10$^5$ |
| Si:Te-1.5% | 0V | 2.4x10$^{-5}$ | 1.2x10$^4$ |
|  | -1V | 5.8x10$^{-7}$ | 5.1x10$^5$ |

## 4. Discussion

In this section, we examine the physical mechanisms that govern the operation of Si:Te PIN photodiodes. It is worth noting that the optically active absorption region in our Si:Te PIN detectors is not the traditional intrinsic Si region but rather the Te-hyperdoped region. Here, the photogeneration of electron-hole pairs with an energy lower than the Si bandgap occurs. To create a steady-state electric field under equilibrium conditions (i.e., with no bias or incident light), we introduce an intrinsic region that is optically transparent to photons with energy lower than 1.1eV. This field arises due to the Coulomb-induced carrier diffusion across the junctions that form in this region. This phenomenon can be expressed by the following equation:

$$\vec{E} = \frac{q \cdot N_D \cdot x_n}{\varepsilon_{Si}} \tag{2}$$

where $\vec{E}$ is the electric field that plays a crucial role in separating and transporting the electron-hole pairs generated by the incident light, $\varepsilon_{Si}$ is the product of the relative permittivity of Si ($\varepsilon_r$) and the permittivity of free space ($\varepsilon_0$). The parameter $x_n$ ($x_p$) is the width of the depletion region, which extends inside the *n-Si:Te* region (*p-Si:B region*), $q$ is the electron charge, while $N_D$ ($N_A$) is the density of active Te dopants (active B dopants) in the *n-Si:Te* region (*p-Si:B* region). The depletion region refers to the full width of the intrinsic region, as well as the extension of the depletion region in the hyperdoped *n*-Si:Te ($x_n$) and *p*-Si:B ($x_p$) regions, which are given by:

$$x_n = \sqrt{\frac{2 \cdot \varepsilon_{Si} \cdot (V_{bi}+V_R)}{q \cdot N_D \left(1+\frac{N_D}{N_A}\right)}}, \quad x_p = \sqrt{\frac{2 \cdot \varepsilon_{Si} \cdot (V_{bi}+V_R)}{q \cdot N_A \left(1+\frac{N_D}{N_A}\right)}} \tag{3}, \quad \text{and } V_{bi} = \phi_T \ln\left(\frac{N_A \cdot N_D}{n_i^2}\right) \tag{4},$$

where $V_R$ is the reverse bias voltage and $V_{bi}$ is the built-in voltage, $\phi_T$ is the thermal voltage $\approx$ 25.9 mV and $n_i$ is the intrinsic carrier concentration.

Focusing on the n-Si:Te side, under unbiased conditions, the length $x_n$ (and $x_p$) is extremely small typically $\leq 10^{-9}$ m due to hyperdoping (using eq. 3). When photons with an energy greater than 1.1 eV impinge on the *n*-Si:Te region (*i.e.,* the active absorber material), the photogeneration of electron-hole pairs occurs. Due to the absence of an electric field in the *n*-Si:Te region, the generated electrons in this region can only be transported through the slower diffusion currents. The holes on the other hand rely on both, the diffusion and the drift currents. They reach the intrinsic region using diffusion current, from where they can be rapidly drifted with high velocities due to the presence of the electric field $\vec{E}$ as shown in Fig. 5 (b).

When a finite reverse bias voltage $V_R$ is applied, most of the voltage drop appears across the high resistance intrinsic region, while minimal voltage appears across *n*-Si:Te (and *p*-Si:B) region. As a result, the depletion region extends deeper (eq. 3 and 4) into the *n*-Si:Te absorber layer from both sides of the *n*-Si:Te structure and can be represented as $x'_n$. This causes the electric field to increase proportionally (using $x'_n$ in eq. 2) and can be represented as $\vec{E}'$. The increase in the electric field enhances the separation of the photogenerated electron-hole pairs in the *n*-Si:Te region and will cause both carriers to drift with a higher velocity. The drift current dominates near the edges of the *n*-Si:Te region marked with $x'_n$ in Fig. 5a, whereas the diffusion current dominates in the center of *n*-Si:Te region (*i.e.* away from $x'_n$ region). The total generated photocurrent density is therefore the sum of drift and diffusion current densities and is given by

$$\vec{J}_{total} = q \cdot \vec{E}(\mu_h \cdot p + \mu_n \cdot n) + q\left(D_h \frac{d\,(p)}{dx} + D_n \frac{d\,(n)}{dx}\right) \quad (5),$$

where $\vec{E}$ is the electric field, $\mu_n$ and $\mu_h$ are the electron and hole mobilities, $n$ and $p$ are the electron and hole concentration respectively. $D_n$ and $D_h$ are the electron and hole diffusion constants. Outside the drift region, only the photons detected within one diffusion length of the depletion region can contribute to a detectable diffusion current. Electron-hole pairs generated further apart from this drift region are more likely to recombine due to the recombination centers formed from the implantation and the PLA-induced defects (in *n*-Si:Te region). The diffusion length for holes is given by

$$L_h = \sqrt{\frac{kT\mu_h\tau_h}{q}} \text{ (conversely } \mu_n \text{ and } \tau_n \text{ for electrons)} \quad (6),$$

Increasing Te doping to 1.5 at.% causes an increase in the absorption coefficient ($\alpha$) of the Te-hyperdoped material, resulting in the generation of more photo-generated electron-hole pairs for the same amount of incident optical power leading to higher responsivity, especially at finite bias voltage (see table 2). For the Si:Te 0.5% PIN detector, the responsivity increased by two orders of magnitude from $1.0\times10^{-7}$ A/W (0V) to $1.2\times10^{-5}$ A/W (-1V), while for the Si:Te 1.5% PIN detector, it increased by three orders of magnitude from $1.1\times10^{-7}$ A/W to $4.2\times10^{-4}$ A/W (-1V) for 1500 nm wavelength. However, the improvement in responsivity for this detector comes at the expense of an increase in the dark current due to the presence of more defects.

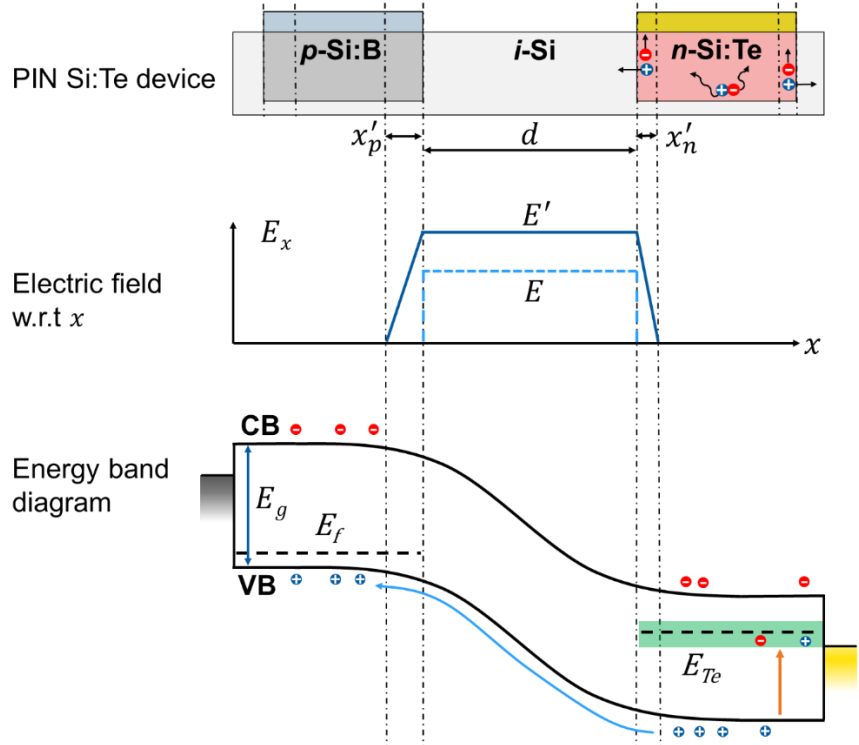

Figure 5. Schematic of PIN Si:Te detector a) depletion region width, b) electric field, and c) energy band diagram upon reverse biasing.

## 5. Conclusion

Our study has demonstrated the successful fabrication of a lateral PIN Te-hyperdoped Si infrared photodetector that operates at room temperature and exhibits increased responsivity with increasing Te concentration. We observed that increasing the Te concentration in the lateral PIN Te-hyperdoped Si infrared photodetector leads to improved infrared absorption. However, this comes at a cost, as we observed that higher Te concentrations introduce more defect states during the fabrication process. These defects are not fully recovered during the PLA process and ultimately result in an enhancement of dark-noise current and reduced carrier mobility, which can negatively impact device performance. Our study also shows that devices with lower doping Te concentrations exhibit higher detectivity and lower NEP. Our findings suggest that the current limitations in device performance can be overcome by optimizing the device dimensions and minimizing the introduction of defects during ion implantation and PLA processes. Specifically, we believe that removing the introduced defects could lead to improved responsivity, lower NEP, and faster response times. By addressing these key factors, we anticipate that the lateral PIN Te-hyperdoped Si infrared photodetector can be further enhanced and optimized for a wide range of sensing and imaging applications.

## 6. Acknowledgements

This work was financially supported by the German Research Foundation (DFG, WA4804/1-1, Projektnummer 445049905). Support from the Ion Beam Center (IBC) and the clean room facility at Helmholtz-Zentrum Dresden-Rossendorf (HZDR) is gratefully acknowledged. M.W. thanks the financial support from the National Natural Science Foundation of China (NSFC) (grant No. 12205212) and the Sichuan Normal University starting founding (XJ20220201)

# References:

1. Lin, H. *et al.* Mid-infrared integrated photonics on silicon: a perspective. *Nanophotonics* **7**, 393–420 (2018).

2. Casalino, M., Coppola, G., Iodice, M., Rendina, I. & Sirleto, L. Near-Infrared Sub-Bandgap All-Silicon Photodetectors: State of the Art and Perspectives. *Sensors* **10**, 10571–10600 (2010).

3. Grote, R. R. *et al.* Extrinsic photodiodes for integrated mid-infrared silicon photonics. *Optica* **1**, 264 (2014).

4. Infrared Detectors and Systems | Wiley. *Wiley.com* https://www.wiley.com/en-us/Infrared+Detectors+and+Systems-p-9780471122098.

5. Marris-Morini, D. *et al.* Germanium-based integrated photonics from near- to mid-infrared applications. *Nanophotonics* **7**, 1781–1793 (2018).

6. Sheng, Z., Liu, L., Brouckaert, J., He, S. & Thourhout, D. V. InGaAs PIN photodetectors integrated on silicon-on-insulator waveguides. *Opt. Express* **18**, 1756–1761 (2010).

7. Yin, D. *et al.* High-responsivity InGaAs/InP photodetectors integrated on silicon-on-insulator waveguide circuits. in *2015 14th International Conference on Optical Communications and Networks (ICOCN)* 1–3 (2015). doi:10.1109/ICOCN.2015.7203674.

8. Chow, P. K., Lim, S. Q., Williams, J. S. & Warrender, J. M. Sub-bandgap photoresponse and leakage current analysis in gold thin film-hyperdoped silicon photodiodes. *Semicond. Sci. Technol.* **37**, 124002 (2022).

9. Pastor, D. *et al.* Interstitial Ti for intermediate band formation in Ti-supersaturated silicon. *J. Appl. Phys.* **112**, 113514 (2012).

10. Liu, F. *et al.* Suppressing the cellular breakdown in silicon supersaturated with titanium. *J. Phys. Appl. Phys.* **49**, 245104 (2016).

11. Dissanayake, S. S. *et al.* Impact of ion implantation and laser processing parameters on carrier lifetimes in gold-hyperdoped silicon. *Semicond. Sci. Technol.* **38**, 024003 (2023).

12. Hu, S. *et al.* Dependence of the optoelectronic properties of selenium-hyperdoped silicon on the annealing temperature. *Mater. Sci. Semicond. Process.* **16**, 987–991 (2013).

13. Kim, T. G., Warrender, J. M. & Aziz, M. J. Strong sub-band-gap infrared absorption in silicon supersaturated with sulfur. *Appl. Phys. Lett.* **88**, 241902 (2006).

14. Lin, A. L., Crouse, A. G., Wendt, J., Campbell, A. G. & Newman, R. Electrical and optical properties of tellurium-doped silicon. *Appl. Phys. Lett.* **38**, 683–685 (1981).

15. Schaub, R., Pensl, G., Schulz, M. & Holm, C. Donor states in tellurium-doped silicon. *Appl. Phys. Solids Surf.* **34**, 215–222 (1984).

16. Lim, S. Q. *et al.* Process-induced defects in Au-hyperdoped Si photodiodes. *J. Appl. Phys.* **126**, 224502 (2019).

17. García-Hemme, E. *et al.* Room-temperature operation of a titanium supersaturated silicon-based infrared photodetector. *Appl. Phys. Lett.* **104**, 211105 (2014).

18. Said, A. J. *et al.* Extended infrared photoresponse and gain in chalcogen-supersaturated silicon photodiodes. *Appl. Phys. Lett.* **99**, 073503 (2011).

19. Wang, M. & Berencén, Y. Room-Temperature Infrared Photoresponse from Ion Beam–Hyperdoped Silicon. *Phys. Status Solidi A* **218**, 2000260 (2021).

20. Wang, M. *et al.* Extended Infrared Photoresponse in Te-Hyperdoped Si at Room Temperature. *Phys. Rev. Appl.* **10**, 024054 (2018).

21. Fang, Y., Armin, A., Meredith, P. & Huang, J. Accurate characterization of next-generation thin-film photodetectors. *Nat. Photonics* **13**, 1–4 (2019).

22. Shaikh, M. S. *et al.* Phase evolution of Te-hyperdoped Si upon furnace annealing. *Appl. Surf. Sci.* **567**, 150755 (2021).

23. Burke, B. G. *et al.* Raman study of Fano interference in p-type doped silicon. *J. Raman Spectrosc.* **41**, 1759–1764 (2010).

24. Fano, U. Effects of Configuration Interaction on Intensities and Phase Shifts. *Phys. Rev.* **124**, 1866–1878 (1961).

25. Yogi, P. *et al.* Interplay between Phonon Confinement and Fano Effect on Raman line shape for semiconductor nanostructures: Analytical study. *Solid State Commun.* **230**, 25–29 (2016).

26. Geis, M. W. *et al.* All silicon infrared photodiodes: photo response and effects of processing temperature. *Opt. Express* **15**, 16886 (2007).

27. Logan, D. F., Jessop, P. E. & Knights, A. P. Modeling Defect Enhanced Detection at 1550 nm in Integrated Silicon Waveguide Photodetectors. *J. Light. Technol.* **27**, 930–937 (2009).

28. Svensson, B. G., Jagadish, C. & Williams, J. S. Generation of point defects in crystalline silicon by MeV heavy ions: Dose rate and temperature dependence. *Phys. Rev. Lett.* **71**, 1860–1863 (1993).

29. Jones, R. C. A Method of Describing the Detectivity of Photoconductive Cells. *Rev. Sci. Instrum.* **24**, 1035–1040 (1953).

# Supporting information

1. *Noise Equivalent Power procedure*

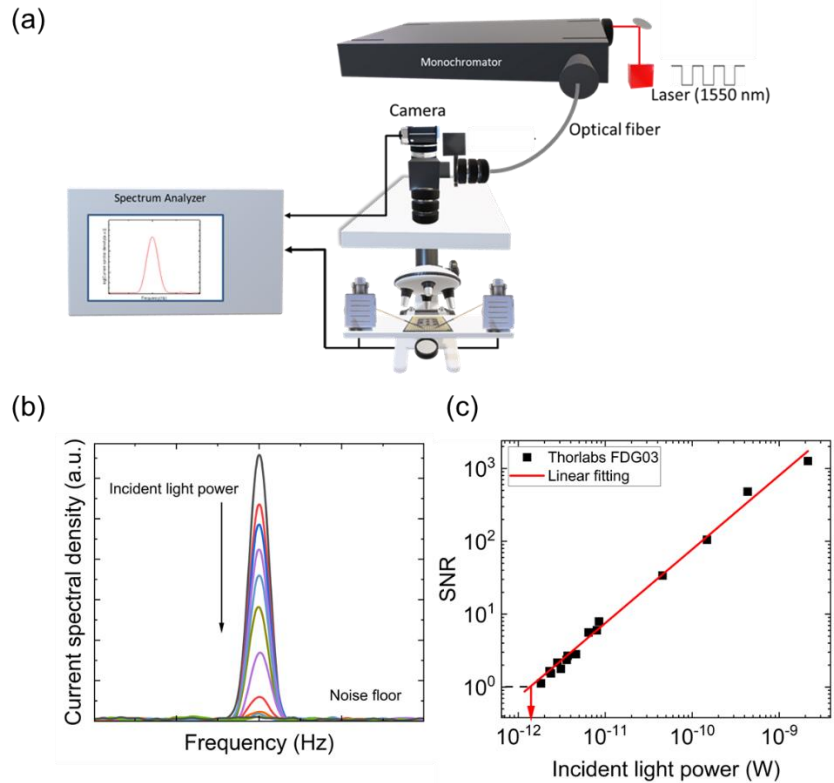

Figure S1. a) Measurement setup to measure NEP b) Noise current spectral density of the first frequency component with attenuated laser intensities, c) SNR plotted against incident light power

For the verification of our measurement procedure, we used a commercially Calibrated Ge detector from Thorlabs with model no. FDG03-CAL, calibrated following DIN EN ISO 9001 and calibration certificate no. 21229134757. We use a modulated laser of 1550 nm wavelength of a certain frequency. The device under test (DUT) is illuminated with this laser, and the generated photocurrent is amplified using a transimpedance amplifier (TIA) with a certain gain and then applied to the spectrum Analyzer. The Spectrum Analyzer converts the time domain signal into a Frequency domain signal, which we need as we are interested in measuring the noise current spectral density. The laser intensity is then attenuated heavily using several neutral Density filters, and the frequency spectrum is measured and saved for each filter. A calibrated photodetector can be used to detect the incident laser power for calculating the responsivity of the DUT.

1. Plot all the noise current spectral densities for the attenuated Laser intensities.
2. Extract the signal spectral peak values at the respective modulation frequency as well as the noise floor value.
3. The signal spectral peak represents the generated photocurrent for a specific laser intensity.
4. Using the responsivity value, estimate the incident optical power by generated photocurrent (A)/Responsivity(A/W) .
5. Plot the SNR vs Incident optical power.

6. The optical power at which the SNR tends to 1, is the noise Equivalent power for FDG03-CAL NEP= 1.2x10$^{-12}$ W/$\sqrt{Hz}$.

7. Calculating specific detectivity $D^* = \frac{\sqrt{A\Delta f}}{NEP}$ where $\Delta f$ is the electrical bandwidth (1 Hz), A is the detector active area 0.0706 cm$^2$, D* is the Specific Detectivity which is equal to 0.2x10$^{12}$cm Hz$^{1/2}$ W$^{-1}$

*2. Control device fabrication*

Si hyperdoped with P to a concentration of 2.5x10$^{20}$ cm$^{-3}$ (0.50 at.%) is prepared as a control device. The doping concentration, as well as the depth, is like the lateral PIN Si:Te devices. The broad amorphization bands after ion implantation can be seen in the as-implanted Si:P material(fig S2 (c)).
B doping in Si did not introduce any noticeable amorphization bands (Fig. S2(d)). The recrystallization of both regions was performed using PLA. The Raman spectra after PLA shows sharp TO modes of Si which represent good crystallinity. The electrical characterization on the Si-doped with P after PLA showed a carrier concentration of 2.2x10$^{20}$ cm$^{-3}$ and a Hall mobility of 56 cm$^2$/Vs, whereas the B-doped side after PLA showed a carrier concentration of 5.1x10$^{19}$ cm$^{-3}$ and a Hall mobility of 42 cm$^2$/Vs. The schematic of the fabricated device is also presented along with the dark IV characteristics (Fig. S2(e-f))

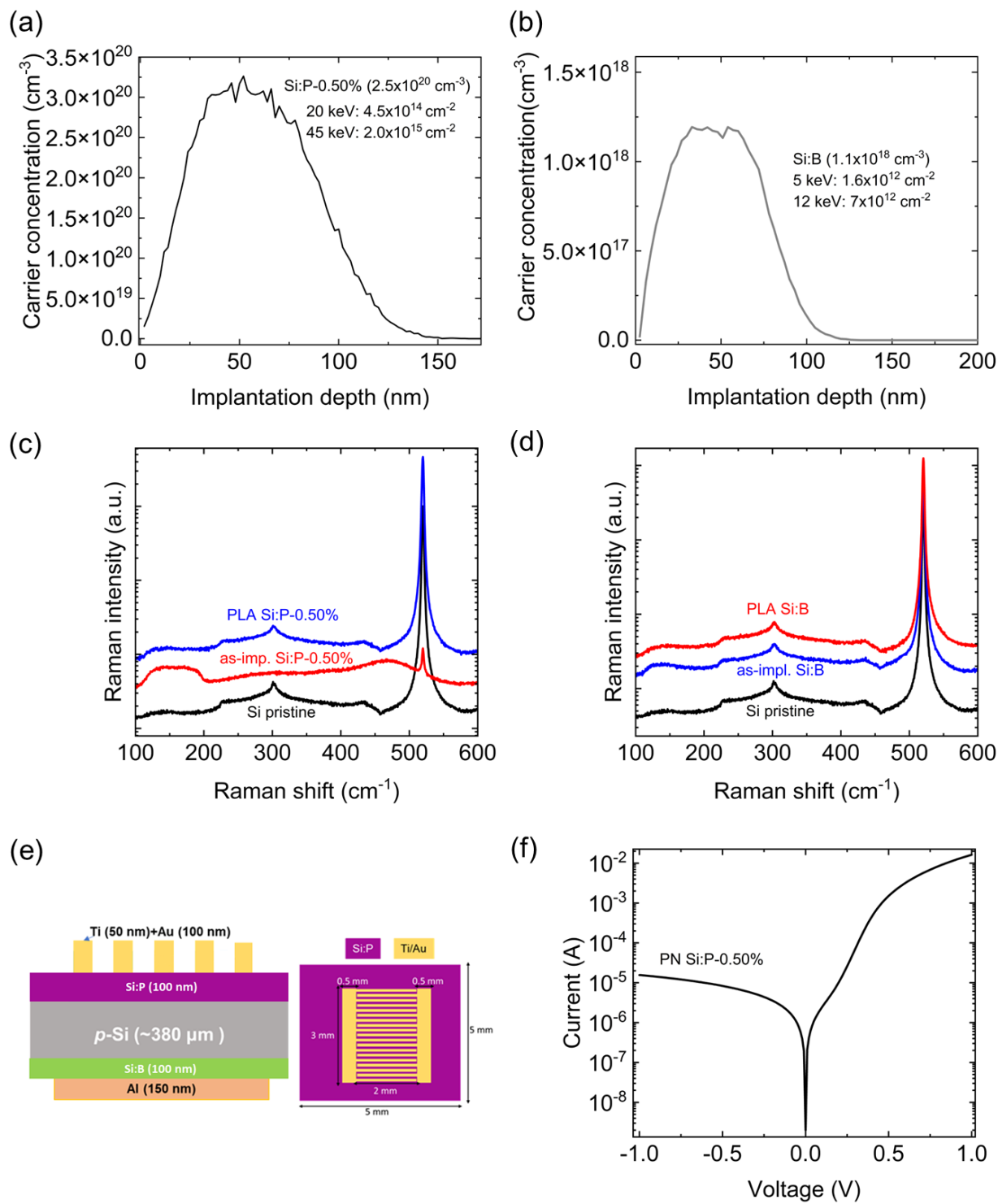

Figure S2. Control device with P hyperdoped absorber region a) device schematic showing cross-section and top view, (b) and (c) SRIM simulation of P and B hyperdoping in Si, (d) Raman analysis of P hyperdoped Si material (e) Raman analysis of B hyperdoped Si material, (f) Dark IV characteristics of Si:P detector.

*3. Raman analysis after PLA*

After pulsed laser annealing of the Te (0.5% and 1.5%) doped silicon material, the 2TA and the TO mode of Si become sharp again representing recovered crystallinity (see Fig. S3). The normalized Raman spectra representing both Te concentration is presented in Fig. S3. Compared to the Si:Te-0.5% sample (in red), the Si:Te-1.5% sample (in blue) shows increased scattering bands around 467 cm$^{-1}$ which originates from higher disorder, induced by ion-implantation which is not completely recovered after the PLA. The Te activation after PLA induces the asymmetric broadening of the TO mode of Si towards the left side of the TO mode (see inset Fig. S3). Compared to the Si pristine sample a Δ FWHM of 0.2 cm$^{-1}$ and 0.7 cm$^{-1}$ is observed for Si:Te-0.5% and Si:Te-1.5% samples respectively. The increased broadening stands for a higher carrier concentration. We also observe softening of the TO Si mode caused by the incorporation of Te substitutional impurities in Si matrix after PLA. The FWHM and peak position of TO mode of Si is summarized in table S1.

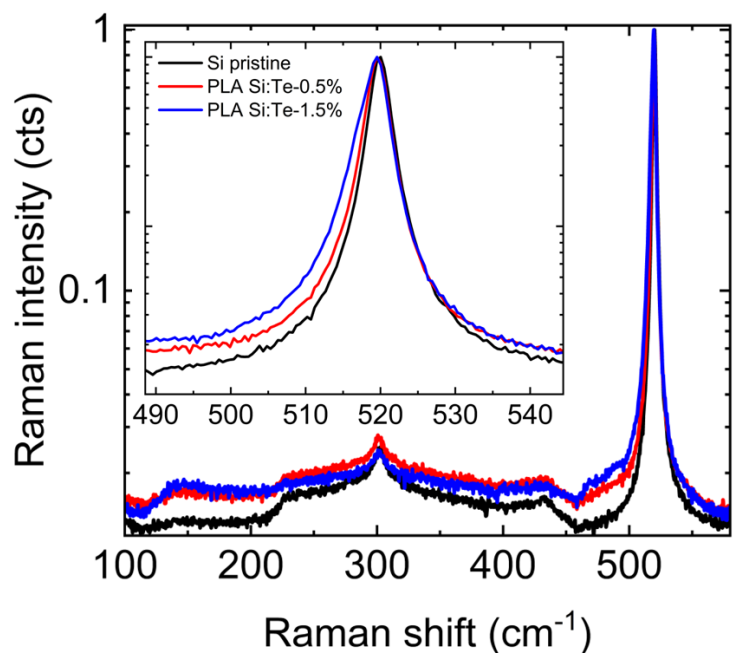

Figure S3. Raman spectra after pulsed laser annealing of Si:Te 0.5% and 1.5% samples.

Table S1. Full width half maximum values and peak position of the Si TO mode after PLA processing of Si:Te samples

| Samples | FWHM (cm$^{-1}$) | Peak position (cm$^{-1}$) |
|---|---|---|
| Si pristine | 2.6 | 519.9 |
| PLA Si:Te-0.5% | 2.8 | 519.5 |
| PLA Si:Te-1.5% | 3.3 | 519.3 |